\documentclass[aps,twocolumn,superscriptaddress,showpacs]{revtex4}

\usepackage{graphicx}
\usepackage{dcolumn}
\usepackage{bm}

\begin{document}


\title{Dynamical formation of stable irregular transients in discontinuous map systems}

\author{Hailin Zou}
\affiliation{Department of Physics and Centre for Computational Science and Engineering, National University of
Singapore, Singapore 117543}

\author{Shuguang Guan}
\affiliation{Institute of Theoretical Physics and Department of
Physics, East China Normal University, Shanghai, 200062, P. R.
China} \affiliation{Temasek Laboratories, National University of
Singapore, Singapore 117508} \affiliation{Beijing-Hong
Kong-Singapore Joint Center of Nonlinear and Complex systems
(Singapore), Singapore 117508}

\author{C.-H. Lai}
\affiliation{Department of Physics and Centre for Computational Science and Engineering, National University of
Singapore, Singapore 117543} \affiliation{Beijing-Hong
Kong-Singapore Joint Center of Nonlinear and Complex systems
(Singapore), Singapore 117508}

\date{\today}

\begin{abstract}
Stable chaos refers to the long irregular transients, with a negative
largest Lyapunov exponent, which is usually observed in certain
high-dimensional dynamical systems. The mechanism
underlying this phenomenon has not been well studied so far. In
this paper, we investigate the dynamical formation of stable
irregular transients in coupled discontinuous map systems.
Interestingly, it is found that the transient dynamics has a hidden
pattern in the phase space: it repeatedly approaches a basin
boundary and then jumps from the bundary to a remote region in the
phase space. This pattern can be clearly visualized by measuring the
distance sequences between the trajectory and the basin boundary.
The dynamical formation of stable chaos originates from the
intersection points of the discontinuous boundaries and their images.
We carry out numerical experiments to
verify this mechanism.
\end{abstract}

\maketitle

\section{Introduction}

Usually, nonlinear dynamical systems may have stable attractors
such as fixed points, limited cycles, or chaotic
attractors. Apart from these, interestingly, it is shown that
irregular long transients can also occur in such systems. For
continuous dynamical systems, these irregular long
transients are usually due to the existence of high-dimensional
chaotic saddles in phase space (see recent review \cite{te lai
review} for details). These chaotic saddles often appear after
crisis bifurcation \cite{crisis}. The system may spend
extremely long time in the vicinity of the chaotic saddle, and
behaves as irregular as chaotic. Due to this reason, this type of
irregular transients is usually called transient chaos, which is
sensitive to the initial condition and thus have a positive largest
Lyapunov exponent. For example, in Ref. \cite{unstable-unstable
pair}, it is found that the development of transient chaos is
related to the unstable-unstable pair bifurcation which involves
an unstable periodic orbit in the chaotic attractor and another
one on the basin boundary. Moreover, another interesting finding
along this line is the super-transient whose average lifetime
could be very long even far from the bifurcation point
\cite{super-transient early,super transient lai}. Such super-
transients have also been found in stochastic dynamical systems
\cite{Extraordinarily-super,do-lai scaling}.

However, there exists another distinct type of irregular transient
that has negative largest Lyapunov exponent, usually occurring in
discontinuous systems. This phenomenon was first observed in the
coupled map lattice \cite{earily_cml}. The complex transients
behave irregularly with exponential decay of correlation both in
time and space \cite{DCA_93}. In addition, the transient time
usually grows exponentially with system size which makes the
attractors unreachable in large systems. Due to these properties,
the transient is termed as stable chaos
\cite{DCA_93,information_flow}.  Later, stable chaos was also
reported in various types of dynamical systems
\cite{phase_transition,stochastic_model,ode_example}. Recently, it
was found in the pulse-coupled oscillators systems which are
frequently used to model neuronal activity
\cite{definition,stable_I_T,09zillmer}. In all the above works,
the stable chaos appears in discontinuous map systems (or discontinuous return maps). Interestingly,
it is found that the stable chaos could also appear in the
continuous map system where there exists a transition from
the standard chaos to the stable chaos \cite{continuous_map}.  

There are some efforts attempting to illustrate the mechanism
underlying the formation of stable chaos. For example, in Ref.
\cite{earily_cml}, it was conjectured that the stable chaos is due to the hierarchical organization of subbasins in phase space. The subbasins are subspaces of a basin separated by walls through which an orbit cannot pass except at portals. The irregular transient is regarded as a sequence of transitions through a hierarchy of subbasins. However, the formation of these subbasins and portals is still not well understood.
In Ref. \cite{Mechanism2}, the stable
chaos was attributed to the ordinary chaos in a continuous system slightly altered from the original discontinuous system.
One deficiency of this approach is that chaos can exist even in a
one-dimensional continuous map, while stable chaos typically
happens in high-dimensional dynamical systems. In addition, Ref.
\cite{information_flow} showed that the alteration could be too
large for some systems. It was shown that the stable chaos is analogous to deterministic
cellular automata \cite{DCA_93}. Along this line, the stable chaos was attributed to the nonlinear propagation of finite disturbances from the outer regions\cite{information_flow}, and a stochastic model was presented to understand the mechanism of this nonlinear information flow \cite{stochastic_model}. However, the direct basic mechanism is still unclear. In spite of the works mentioned above, the dynamical formation of stable chaos has not been well
studied to date and the mechanism is still unclear. In particular, how the stable chaos develops from the governing dynamical equations is not fully understood. One possible reason
for this is that these global behaviors usually occur in
high-dimensional dynamical systems, which is usually difficult to
deal with mathematically.

In this paper, we focus on the problem on the dynamical formation of stable chaos in the discontinuous systems. For
these systems, it is natural to relate the
occurrence of stable chaos to the discontinuity of the local
dynamics of the coupled dynamical systems. Motivated by this idea,
in this paper, we directly investigate some coupled discontinuous
map systems, which are different from the previous works. Our
particular interest is to reveal how the discontinuity in the
local dynamics of a coupled system can induce stable chaos. To
this end, we specifically construct dynamical models whose local maps
have only contracting pieces with absolute determinant smaller than $1$. As a consequence, the occurrence of
chaos is excluded in such systems, and the generated long irregular
transient is stable chaos by nature.

The organization of this papers is as follows. In Sec. II, we show
that accompanying stable chaos, a regular pattern always exists in
both coupled discontinuous maps and pulse coupled oscillators. In
Sec. III, the dynamical origin of stable chaos is analyzed and
verified by a two-dimensional map. Finally, conclusions are
drawn in the last section.

\section{The regular pattern accompanying stable chaos}

In continuous dynamical systems, the stable manifolds of the
saddle periodic orbits compose the basin boundaries. However, for
discontinuous map systems with only contracting local
dynamics, the basin boundaries could only include the set of
points whose dynamics are discontinuous. This set of points
comprises the pre-images of the discontinuous boundaries and the
discontinuous boundaries themselves. Intuitively, the stable chaos
might be related to this set. Normally, we cannot obtain this set
directly because of the high dimensionality of the phase space.
However, the basin boundaries can be easily detected.
We can measure the distance of each  point in the trajectory
to the basin boundaries. Therefore, for each transient trajectory, we will
have a corresponding distance sequence.

The distance of a point $x$ to the basin boundaries $B$ is defined to
be $d=\mathrm{min}\left\Vert x-y\right\Vert _{2}$, where $y \in B$.  
This distance $d(x)$ also quantifies the degree of stability of
the system under finite perturbation at a given point $x$. If the
perturbation added to the system  is larger than $d(x)$, the
system will jump to another attractor. In this sense, the distance
is the maximal finite perturbation that the system can tolerate
without losing the stability. 
We use a simple procedure to obtain this distance. The steps are:
(1) Randomly sprinkle many initial points, for example 200, in a hypersphere with center $x$ and radius $r$. Initially $r$ is set to be a large value. If all the initial points settle onto the same attractor as the center $x$, go to step (3). (2) Set the new radius $r$ to be the minimum distance between the center and the set of initial points who have different final attractor from the center. (3) If $r$ does not change in the sequential $m$, say 5, times, stop. Otherwise repeat the above steps.

To illustrate this idea, we first consider the following coupled map lattice with a periodic
boundary \cite{earily_cml}:
\begin{equation}\label{eq1}
x_{i}(n+1)=\frac{1}{2r+1}\sum_{j=-r}^{r}f(x_{i+j}(n)),
\end{equation}
where the local dynamics is $f(x)\cong sx+\omega\ (\mathrm{mod}\
1)$. When $s$ is smaller than 1, the system (1) will finally
approach a periodic attractor because of the negative Lyapunov
spectrum. We choose $s=0.9,$ and $\omega=0.118$ in this paper. We consider nearest-neighbor coupling: $r=1$. The
number $N$ of oscillators is 28, for which the irregular long
transient is prominent. For the individual local map, there is a
discontinuous boundary at $x=(1-\omega)/s=0.98$.

\begin{figure}[htbp]
\includegraphics[scale=0.25]{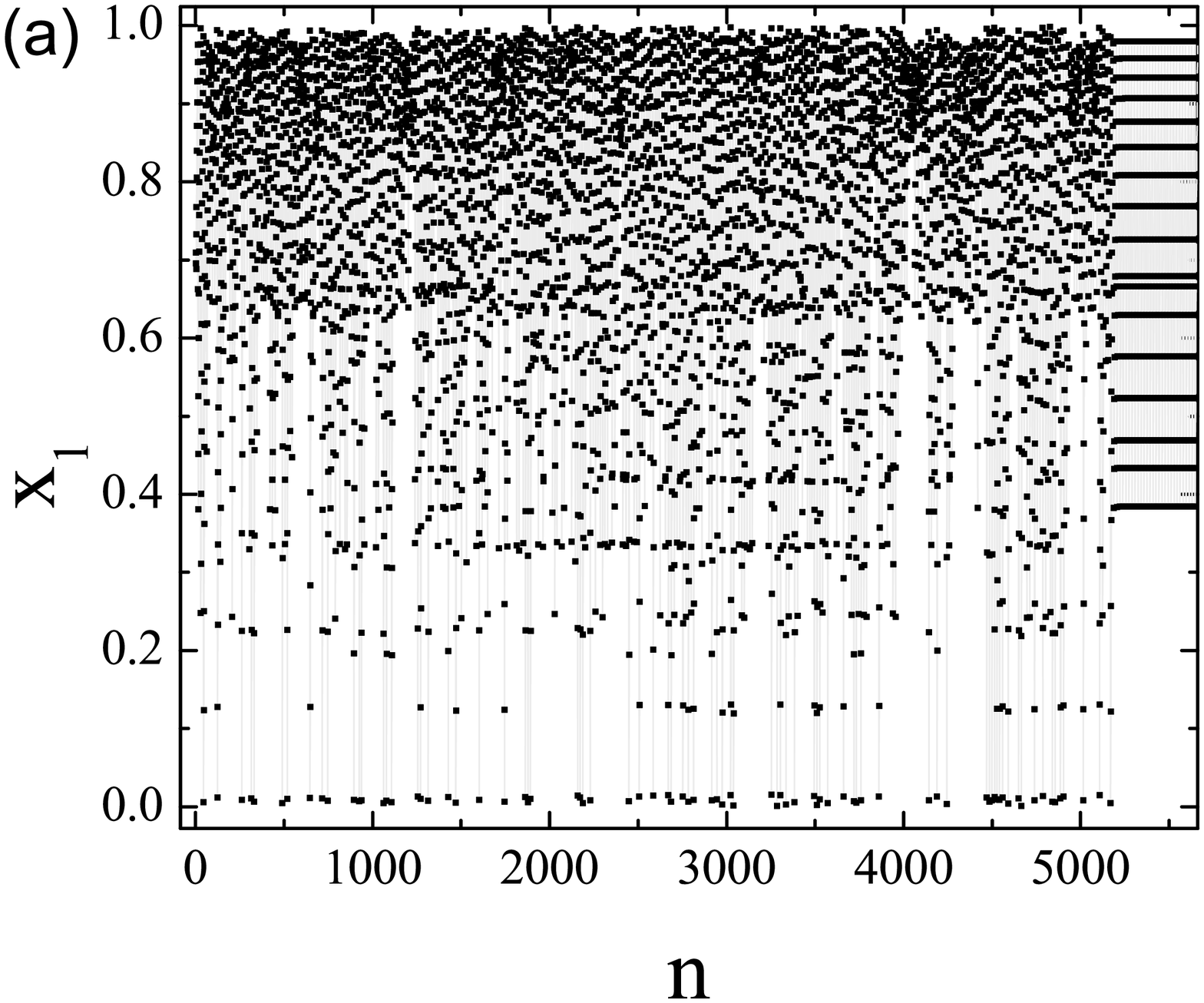}
\includegraphics[scale=0.25]{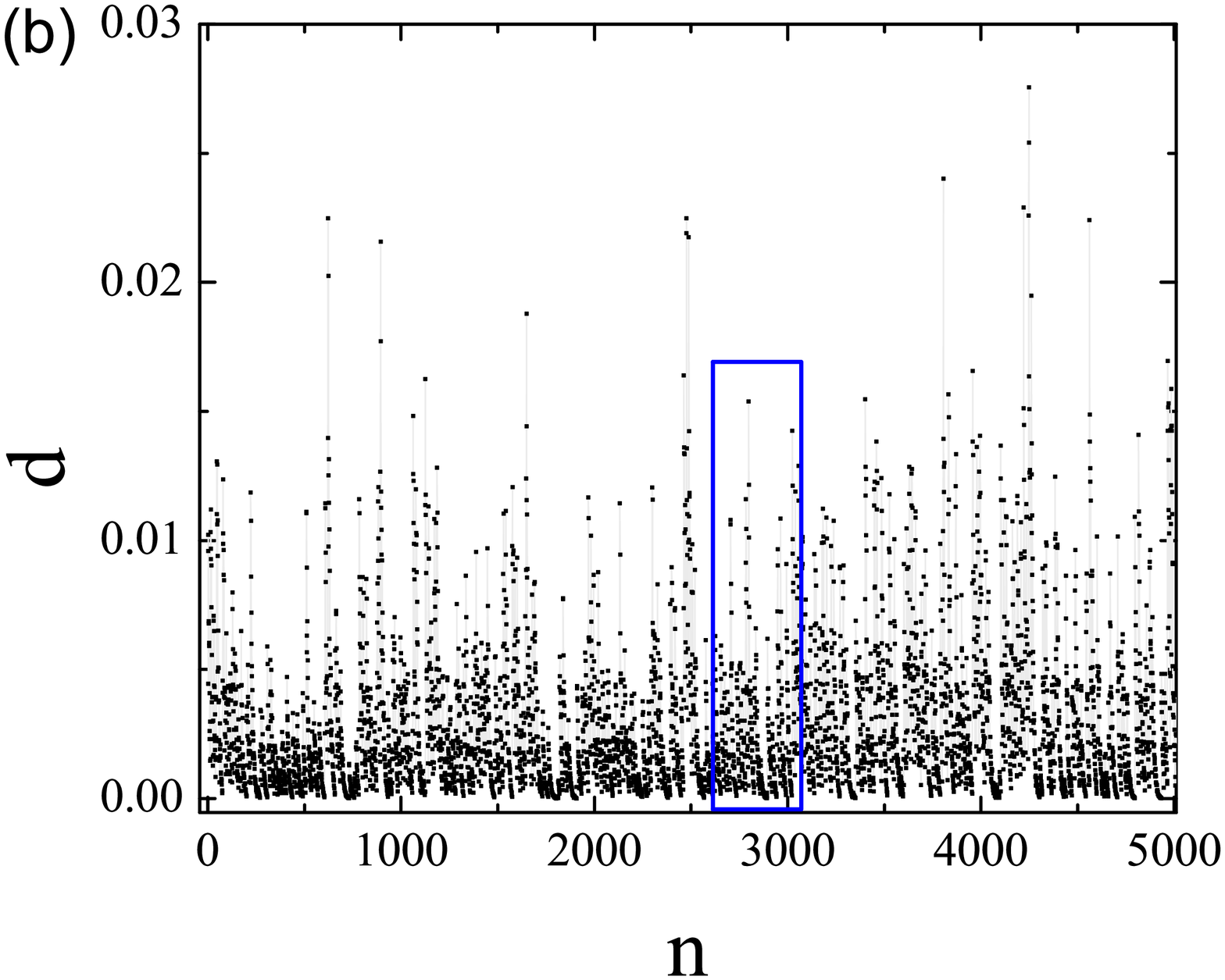}
\includegraphics[scale=0.25]{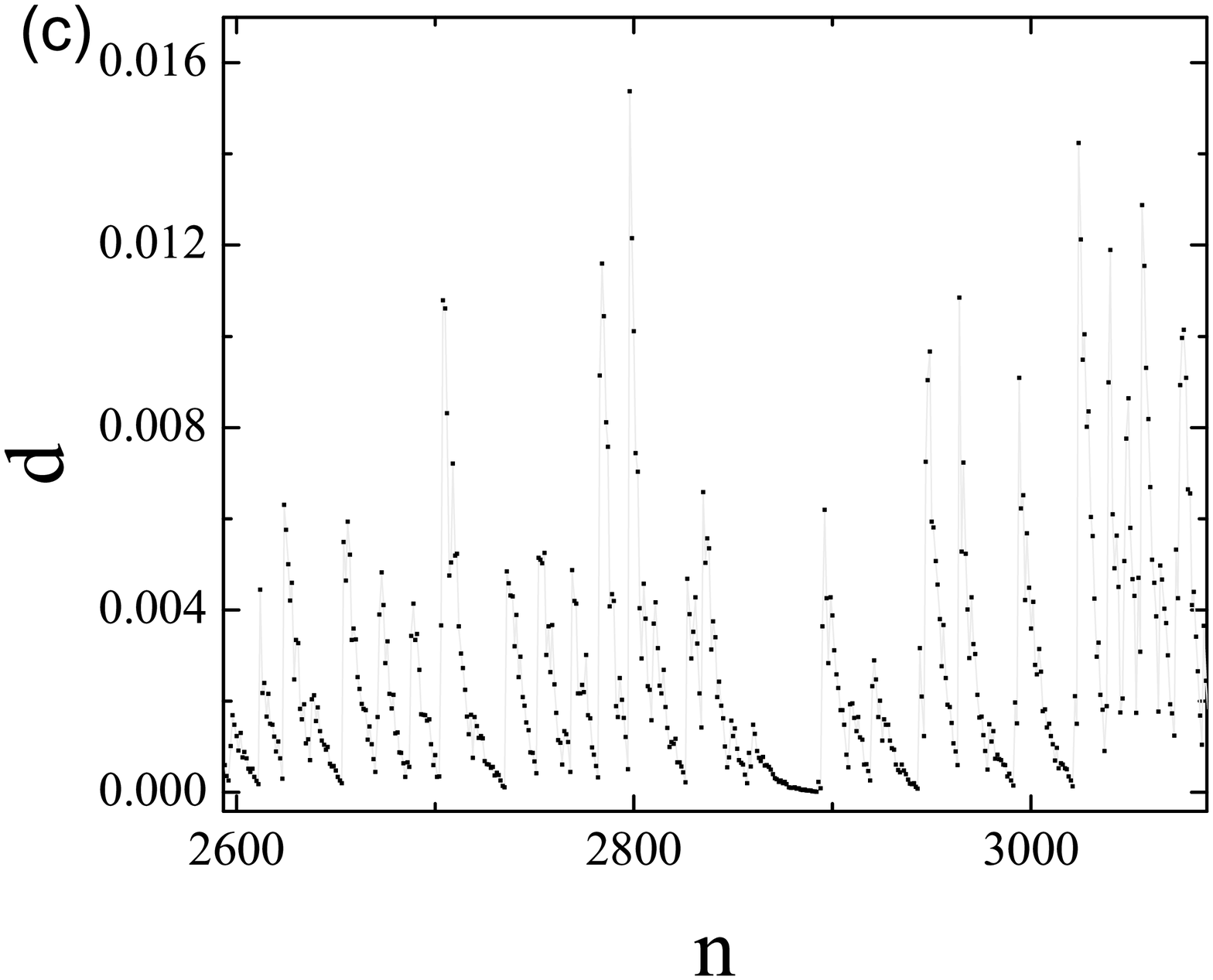}
\caption{(a) The long irregular transients observed in Eq.
(\ref{eq1}) for $x_1$. $n$ is the time step. (b) The corresponding distance
sequence $d$ to the basin boundaries. (c) The enlargement of the
rectangle in (b).}
\end{figure}

Fig. 1(a) shows a typical stable irregular transient in this
system. Its corresponding distance sequence to the basin
boundaries is shown in Figs. 1(b) and 1(c). To our surprise, we
find a regular pattern in the distance sequence despite of the
irregular transient dynamics. As clearly shown in Fig. 1(c), the
distance sequence first gradually approaches the basin boundaries.
Then it jumps to some remote regions in the phase space and begin to
approach the basin boundaries once again. During the whole
transient period, such pattern repeatedly occurs until finally the
system settles down on the trivial attractors.

Similarly, the regular pattern in the distance sequence can also
be observed in the inhibitory pulse-coupled oscillators. This model describes $N$
oscillators, such as neurons, interacting on a direct network by sending and
receiving pulses\cite{MS-model, 95-ernst, stable_I_T,return_map}.
The state of each oscillator $i$ is specified by a phase-like variable
$\phi_{i}(t)\in(-\infty,1]$. The dynamics of the single oscillator $i$ is given by
\begin{equation}\label{eq2}
d\phi_{i}/dt=1.
\end{equation}
when $\phi_{i}(t)$ reaches a phase threshold $1$, this phase is reset
to zero, $\phi_{i}(t^{+})=0$, and a pulse is generated. After a
delay time $\tau$ this pulse is received by all oscillator $j$ having an in-link from $i$. This induces a phase jump in $j$ according to
\begin{equation}
\phi_{j}(t+\tau)^{+}=\min\{U^{-1}(U(\phi_{j}(t+\tau))+\epsilon_{ji}),1\},
\end{equation}
where the function $U$, which determines the phase jump, is twice
continuously differentiable, monotonically increasing, concave and
normalized ($U(0)=0$ and $U(1)=1$). Furthermore, the coupling
strength is also normalized as
$\epsilon_{ij}=\epsilon/k_{j}$,where $k_{j}$ represents the number
of incoming links of node $j$ (in-degree in graph term). This
model is equivalent to the standard leaky integrate-and-fire model
with $U_{i}(\phi)=\gamma_{i}^{-1}I_{i}(1-\exp(-\gamma_{i}\phi))$   \cite{stable_I_T}.

It is convenient to investigate the dynamics in a return map by choosing an arbitrary oscillator as reference\cite{return_map}. Here the oscillator 1 is chosen. When the reference oscillator is reset, the phases are recorded. The number of reset times is used as time step for the return map. The dynamics can be simulated by an event-by-event based numerical
calculation which can be applied to obtain the solutions for the
pulse-coupled network with delay \cite{Event-simulation}.

\begin{figure}[htbp]
\includegraphics[scale=0.25]{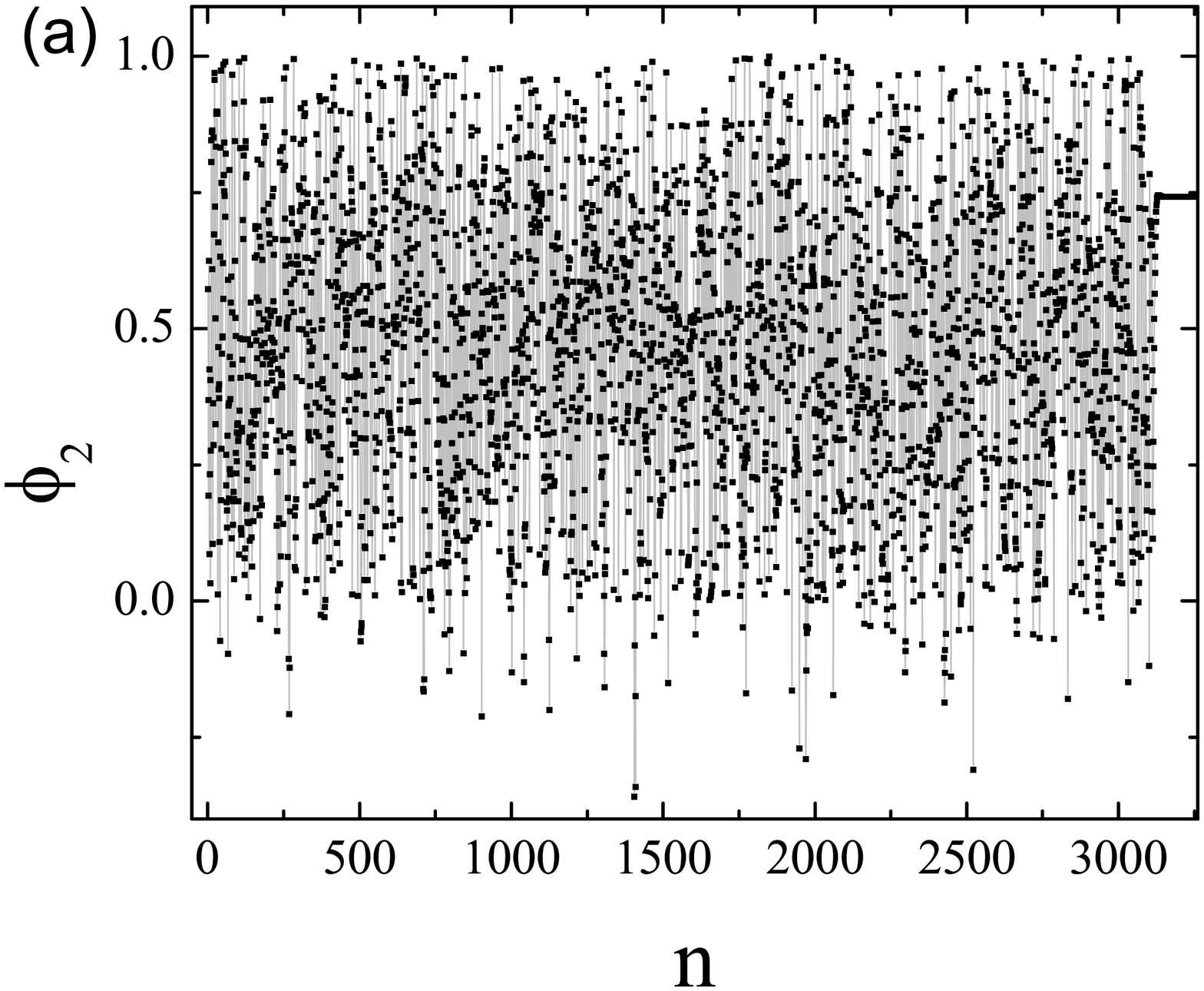}
\includegraphics[scale=0.25]{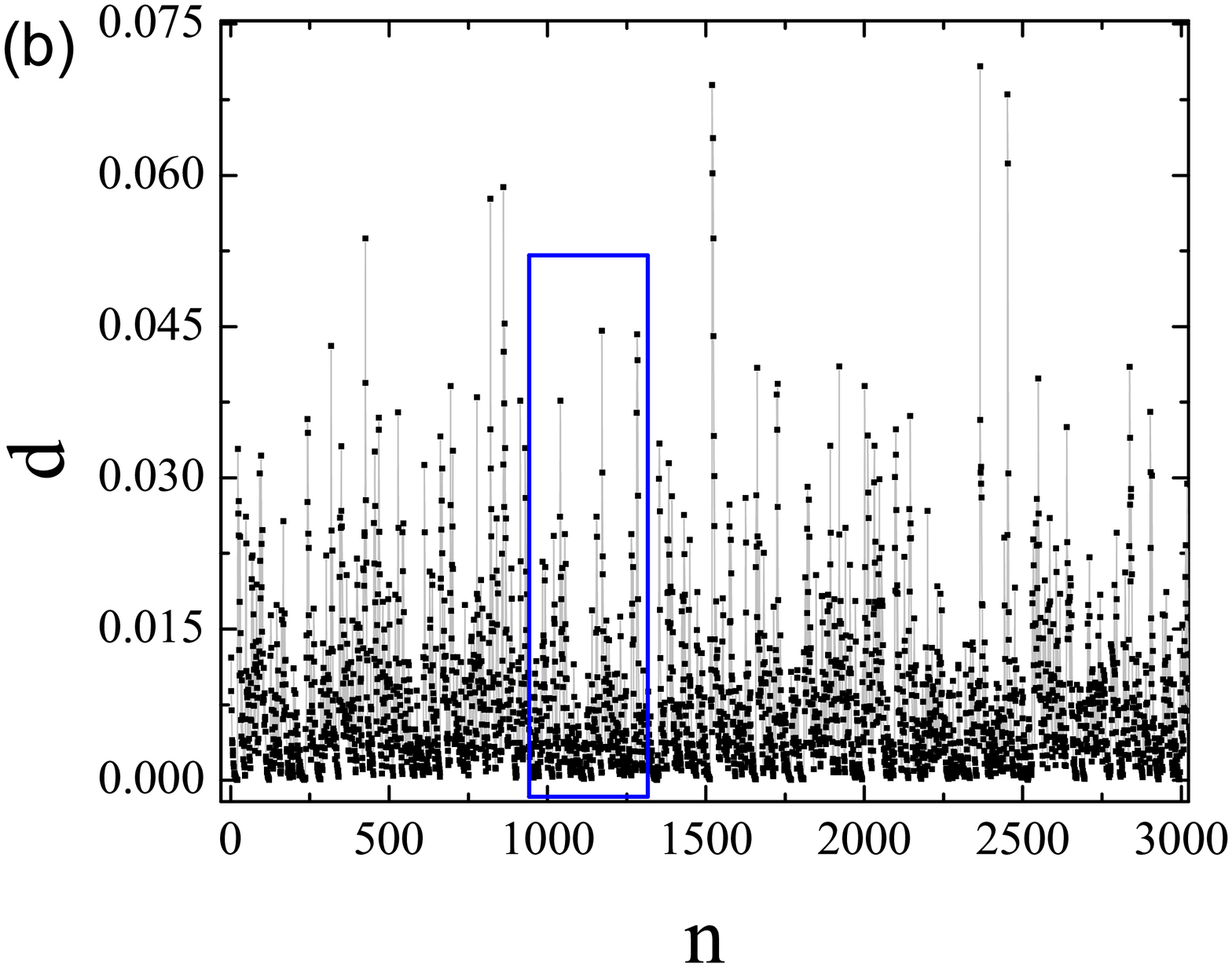}
\includegraphics[scale=0.25]{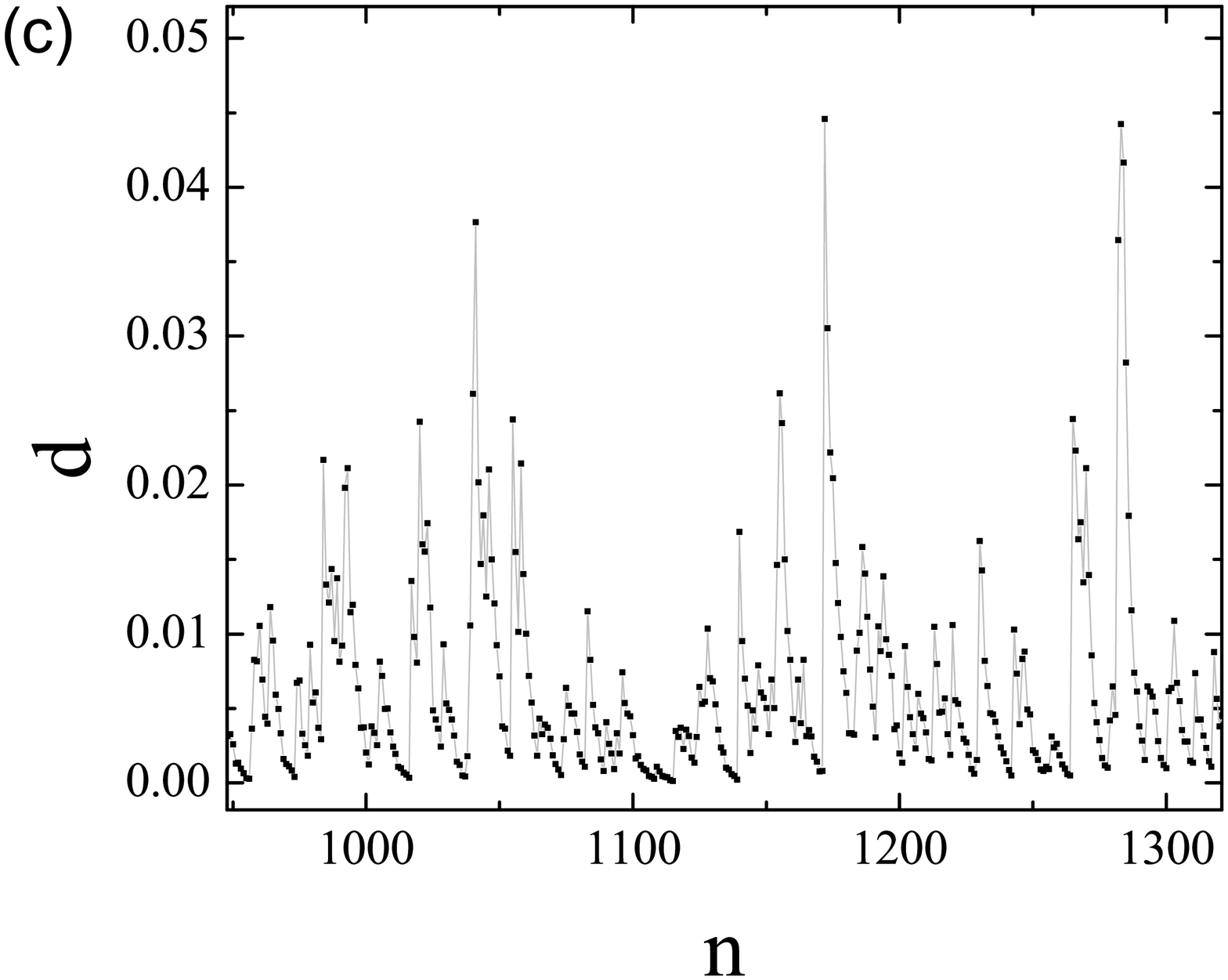}
\caption{(a) The long irregular transients observed for pulse-coupled oscillators for $\phi_{2}$ using oscillator $1$ as reference. The $n$ represents the number of times the reference oscillator has been reset. (b) The corresponding distance
sequence $d$ to the basin boundaries. (c) The enlargement of the
rectangle in (b).}
\end{figure}

We choose the similar parameter values as in Ref.
\cite{stable_I_T}, where  the existence of stable irregular
transients is analytically proved, i.e., $\gamma_{i}=1$,
$I_{i}=4.0$, $b=1$, $\epsilon=-1.6$, and $\tau=0.1$. The network
size is fixed at $N=20$. We observed that long irregular
transients exist when diluting links from fully coupled networks.
Fig. 2(a) shows one typical transient trajectory for connection
probability $p=0.2$. From the distance sequence to the basin
boundaries, as shown in Figs. 2(b) and 2(c), once again, we find
the similar pattern as shown in Fig. 1. During transients, the
trajectory repeatedly goes to the basin boundaries and then jumps
to some remote regions in the phase space. The existence of the same
regular pattern in the above two different systems strongly
suggests that the formation of stable chaos in different discontinuous dynamical
systems actually could have the same dynamical origin. We
emphasize that such pattern is unique in stable chaos, and does
not exist in chaotic transients observed in the excitory
pulse-coupled oscillators \cite{chaotical-transient-pulse}, which
should have a different mechanism.

\section{Dynamical formation of stable chaos}

In Sec. II, we observe a regular pattern in the stable chaos by
measuring the distance to the basin boundaries for each point in
the transient trajectory. This regular pattern displays the
internal structure of the stable irregular transients. During the
transient period, the trajectory repeatedly goes to different
regions on the basin boundaries. This seems to suggest that there
is a path that can connect two different places on the basin
boundaries. Through extensive numerical experiments, we confirm
the existence of such paths.
Furthermore, it is found that the
ending point of such a path is an image of the starting point, and
both the starting point and the ending point are on the
discontinuous boundaries. We thus call this path a {\em guiding
path}. The dynamical formation of this path is through the
intersection of discontinuous boundaries with their images. It can
be easily verified that the transient structure in the system
(\ref{eq1}) is formed in this way where the discontinuous
boundaries are $x_{i}=0.98$. The dynamical origin of stable chaos
lies in the formation of many {\em guiding pathes} in phase space.
At the starting and the ending points, the dynamics are
discontinuous. Between these two points, the trajectory
usually follows the contracting dynamics which results in the
decreasing distance to the basin boundaries.

 To verify the above mechanism of the formation of stable chaos,
 we need to locate guiding paths and then compare with the corresponding
 jumping processes from one region to another region on the basin boundaries.
  For high dimensional systems, such as Eq. (\ref{eq1}), we can obtain
  approximately the underlying guiding path by sampling many initial points,
  say $10^6$, in a small region where
  a jumping process starts. The length of guiding path $n$ usually is the
  same as the jumping process. We can obtain all these trajectories with length $n$. Then we approximate the guiding path by a trajectory with minimum quantity $D$, where $D$ is defined to be the sum of distance of two endpoints of a trajectory to the discontinuous boundaries. In this way, we can verify that there is a guiding path
 governing the jumping process from one region of basin boundaries to another
 region.

 To be more accurate and more reliable, here we construct a two-dimensional
 map and develop a numerical technique to obtain the guiding paths
 with high accuracy. The dynamical equations of the discontinuous maps are:
\begin{eqnarray}\label{eq2d}
x_{1} &=& ef(x_{1})+(1-e)f(x_{2}), \nonumber \\
x_{2} &=& ef(x_{2})+(1-e)f(x_{1}),
\end{eqnarray}
where the local map $f$ is shown in Fig. 3, which can be regarded
as the generalized map used in Eq. (\ref{eq1}) with more
discontinuous boundaries.

 In the following, we specifically describe the technique to locate
 the guiding paths accurately for
 the two-dimensional maps $F$ whose discontinuous boundaries are composed of
 straight lines. The key procedure of this technique
 is to locate the guiding
 paths with given length $n$ and an initial interval $[A,B]$.
 By varying $n$ and the interval, we can locate all the
 guiding paths of interest. To this end, we divide the phase space into squares,
 or {\em cells}
 by the discontinuous boundaries. An interval $[A,B]$ may contain
 starting points of the guiding paths if $F^n(A)$ and $F^n(B)$ fall into
 two different cells. Consider the midpoint $m=(A+B)/2$, we can find a smaller interval $[A,m]$ or $[m,B]$ whose $n$th images of the two endpoints fall into two different cells. We can recursively apply
 this bisection method to shrink the interval and obtain an interval $[a,b]$ with given accuracy,
 say, $10^{-6}$. Then one $n$th image of the two endpoints will be the ending point of the guiding path if it is within the distance $10^{-6}$ of discontinuous boundaries. After that, we consider the remaining interval $[b,B]$. In this way, we can find all guiding paths with length $n$ with starting points belonging to $[A,B]$.

%
\begin{figure}[htbp]
\includegraphics[scale=0.25]{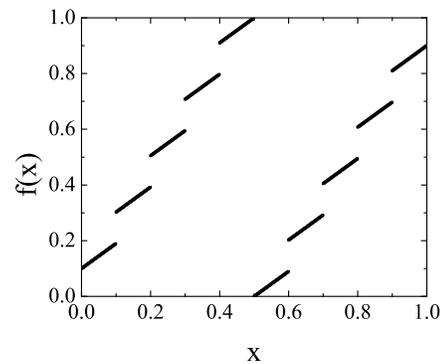}
\caption{ A map with only contracting pieces and many discontinuous boundaries.}
\end{figure}

A typical transient trajectory in  Eq.
(\ref{eq2d})  is shown in Fig. 4(a) for $e=0.09$. In the distance
sequence, as shown in Fig. 4(b), a regular pattern similar to
those in Figs. 1(c) and 2(c) is much more evident. Here, since our
system is only two-dimensional, it is easy for us to directly
verify our analysis of the mechanism that led to the formation of stable chaos. In Fig.
5, we plot part of the transient trajectory within a specific time
window, i.e., from $n=18$ to $n=35$. As shown in Fig. 4(c), during
this time period, the transient trajectory goes from one region of
the basin boundary to another region of the basin boundary. In Fig. 5,
we plot both the transient trajectory from $n=18$ to $n=35$, and a
{\em guiding path}. Remarkably, it is clearly seen that the
transient trajectory exactly follows the {\em guiding path} going
from one point of the basin boundary to another point on the basin
boundary. Since there exist plenty of these {\em guiding pathes}
in the phase space, we can expect long irregular transients to occur
in such coupled map systems, especially when the dimension of the
coupled system is very high.

\begin{figure}[htbp]
\includegraphics[scale=0.25]{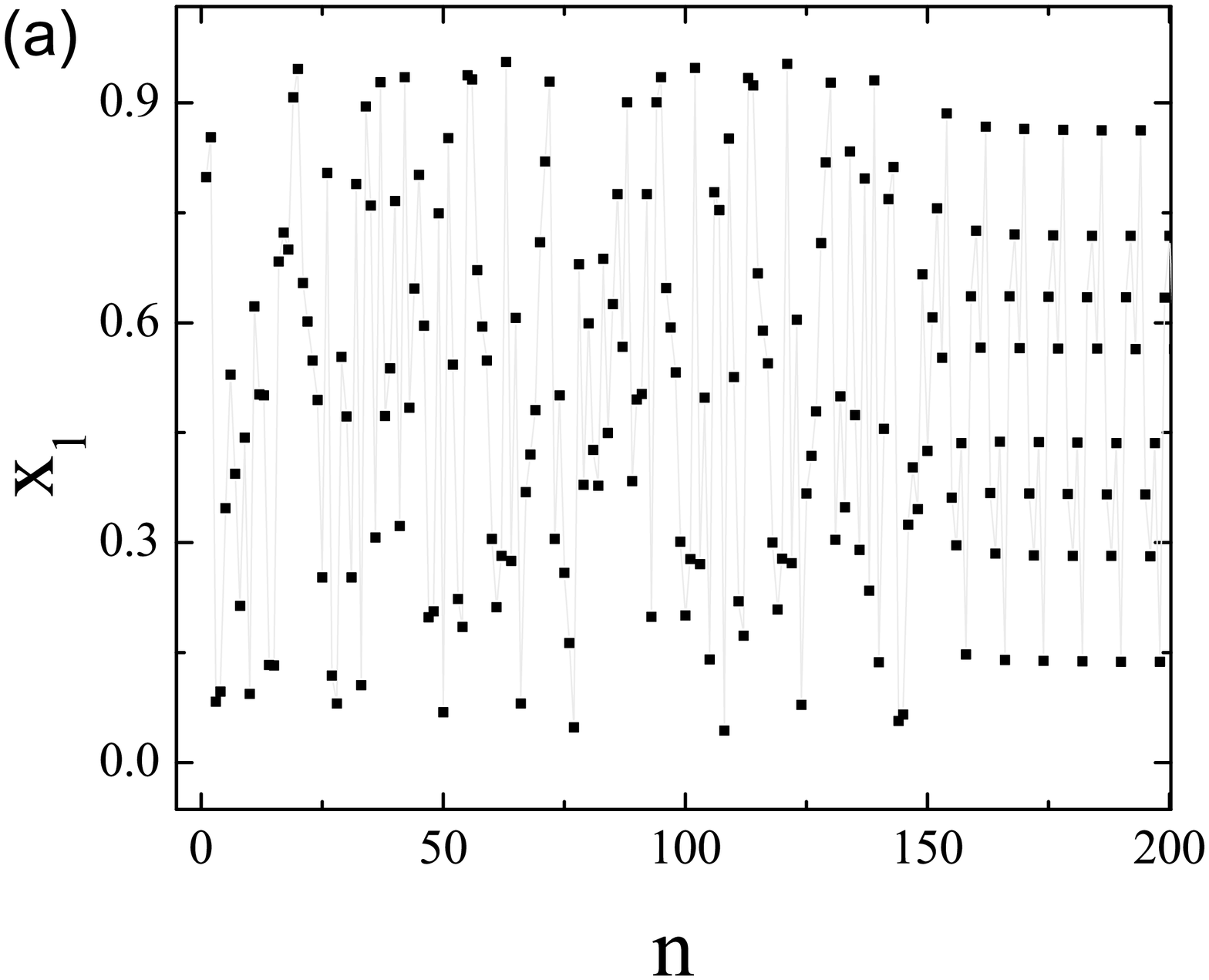}
\includegraphics[scale=0.25]{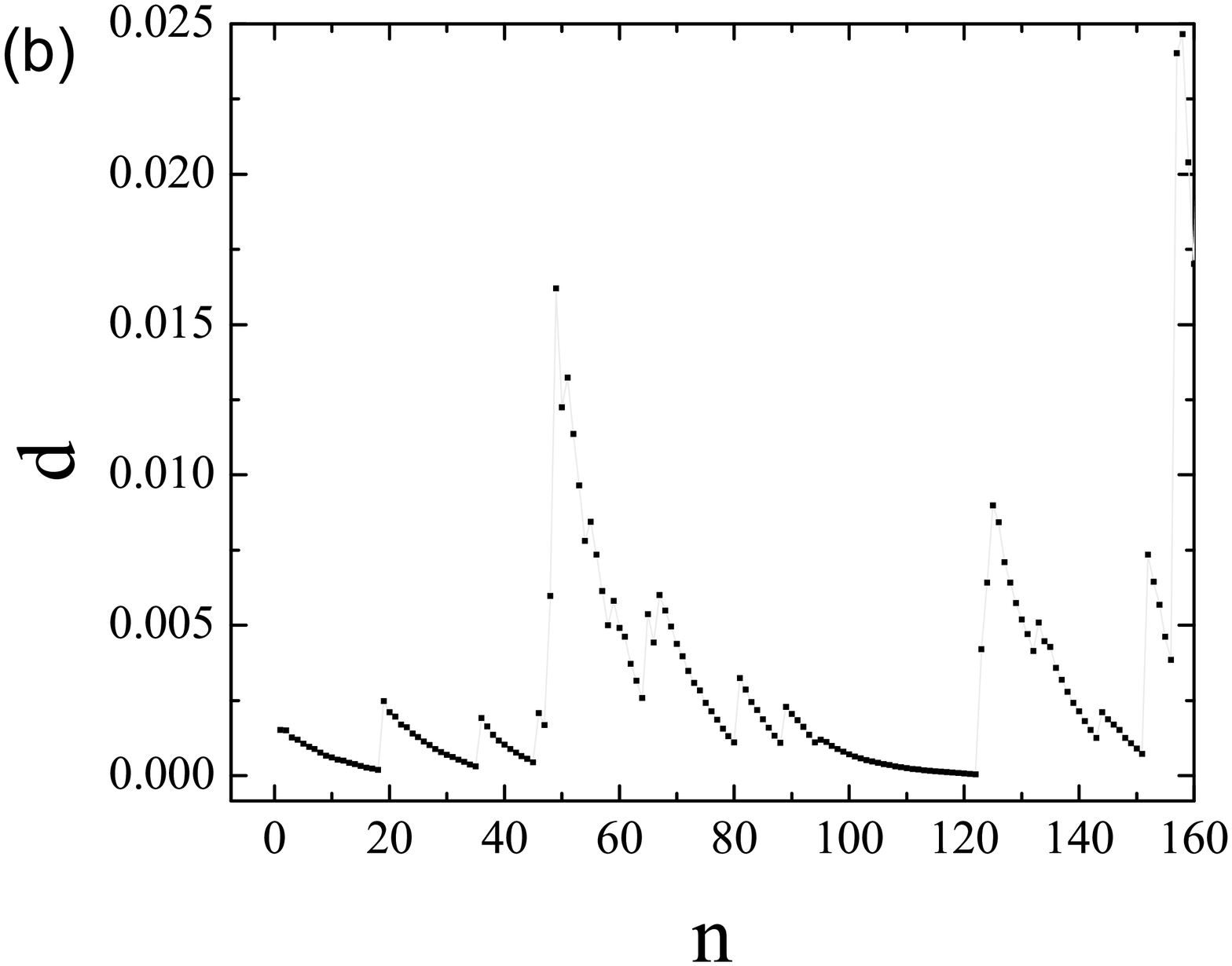}
\caption{A typical transient trajectory (a) for Eq. (\ref{eq2})
and its corresponding distance sequence (b).}
\end{figure}

It is interesting to compare stable chaos to the transient chaos usually occurring in continuous dynamical systems. The skeleton for transient chaos is the infinite number of unstable periodic orbits (UPOs) embedded in the chaotic saddle. While for the stable irregular transient or stable chaos, the underlying microscopic structure is the {\em guiding path} with both starting and ending points on the discontinuous boundaries. A {\em guiding path} is not a cyclic structure, i.e. the starting and ending points are not the same, which is the major difference with an UPO. The regular patterns occurring in the high dimensional systems  Eq. (\ref{eq1}) and Eq. (\ref{eq2}) impliy that the {\em guiding paths} are clustered in phase space, which is similar to the dense UPOs. Here, the clustered {\em guiding paths} are generally associated with the large number of discontinuous boundaries in the high dimensional systems. For Eq. (\ref{eq1}), the number of discontinuous boundaries is $N+{N \choose 2} + {N \choose 3} + {\cdots} $. It is expected that this fast growing of the number of discontinuous boundaries with dimension will make the {\em guiding paths} more clustered in high dimensional systems, somewhat similar to the attractor crowding effect \cite{attractor_crowding}. In turn, it will make stable chaos more easily observed in high dimensional systems.

\begin{figure}[htbp]
\includegraphics[scale=0.25]{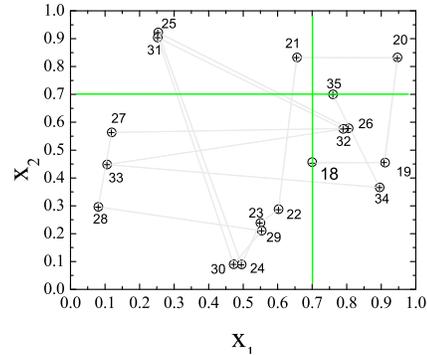}
\caption{Part of transient trajectory shown in plus and a guiding
path shown in circle from $n=18$ to $n=35$. The two lines
represent two discontinuous boundaries. The number represents the
the time step $n$.}
\end{figure}
From the microscopic structure, we can also understand why stable chaos is unstable against finite small perturbation (stable against infinitesimally small perturbation) \cite{09zillmer}. During the long transient time, stable chaos takes places near many {\em guiding paths} whose starting and ending points belong to the basin boundaries.

\section{Conclusion}

In this paper, we investigate the dynamical formation of  long
irregular transients with negative largest Lyapunov exponent,
i.e., the stable chaos, directly based on the dynamical
equations. We show that these irregular transients actually have
certain structure which can be illustrated by the distance
sequence to the basin boundaries. The transients repeatedly
approaches the basin boundaries and then jumps from the boundaries
to a remote region in the phase space. Through numerical simulations, it is shown that there exists a {\em
guiding path} whose ending point is an image of the starting point
and both of them are on the discontinuous boundaries.  It is these
{\em guiding pathes} that connect different points on the basin
boundaries, making the dynamics of the system exhibit long
irregular behavior before it goes to the final stable attractor.
Thus the present work reveals a mechanism for the formation of
stable chaos in coupled discontinuous map systems.

\bigskip
\acknowledgments This work is supported by the National University of
Singapore, and DSTA of Singapore under Project Agreement POD0613356.

\end{document}